\documentclass[12pt]{article}
\usepackage{cjp}

\def\nabstar#1{\nabla\kern-0.5pt\smash{\raise 4.5pt\hbox{$\ast$}}
               \kern-4.5pt_{#1}}

\def\drvstar#1{\partial\kern-0.5pt\smash{\raise 4.5pt\hbox{$\ast$}}
               \kern-5.0pt_{#1}}

\def\newline{\relax\ifhmode\null\hfil\break\else\nonhmodeerr@\newline\fi}
\def\frac#1#2{{#1\over#2}}
\def\text#1{{\hbox{\rm #1}}}
\def\flushpar{{\par \noindent}}

\def\beq{\begin{equation}}
\def\eeq{\end{equation}}
\def\bea{\begin{eqnarray}}
\def\eea{\end{eqnarray}}
\def\Id{ \mbox{1\hspace{-1.2mm}I} }
\def\BE{\begin{equation}}
\def\EE{\end{equation}}
\def\BA{\begin{eqnarray}}
\def\EA{\end{eqnarray}}
\def\BAN{\begin{eqnarray*}}
\def\EAN{\end{eqnarray*}}

\def\Tr{\mathrm{Tr}}
\def\tr{\mathrm{tr}}

\def\gm5{\gamma_5}

\def\det{\mathrm{det}}

\def\CT{{\cal T}}

\def\anxc{{\cal A}_D (x)}

\input epsf.sty
\newdimen\psfigsize
\def\psfigure#1 #2 #3 #4 #5{
    \begin{figure}[tbh]
      \begin{center}
      \vbox{
        \null\vskip-0.2in\hskip#2
        \epsfxsize=#1
        \epsfbox{#4}
        \vskip -0.3in
        \caption {#5 \label{#3}}
        \vskip 0.0 true in plus 0.3 true in
      }
      \end{center}
   \end{figure}
}

\begin{document}

\title{Chiral fermions on a finite lattice\footnote{Talk given at Chiral'99,
"Workshop on Chiral Gauge Theories", Sep. 13-18, 1999, Taipei.} }
\author{Ting-Wai Chiu}
\address{Physics department, National Taiwan University \\
         Taipei, Taiwan 106, Republic of China. }
\date{\today}
\maketitle

\begin{abstract}

We discuss how to formulate Dirac fermion
operator on a finite lattice such that it can provide a
nonperturbative regularization for massless fermion interacting
with a background gauge field.

\end{abstract}


\section{ Introduction }

Consider a massless Dirac fermion interacting with a background
gauge field. Our present goal is to formulate a nonperturbatively
regularized quantum field theory which at least satisfies
the following physical constraints :
\begin{description}
\item{(A)} In the classical limit, it reproduces the
           classical physics of the action, \newline
           $ { \cal A } = \int_x \bar\psi(x)
              \gamma_\mu [ \partial_\mu + i g A_\mu(x) ] \psi(x) $.
\item{(B)} For topologically trivial gauge backgrounds,
           and in the weak coupling limit, it agrees with the predictions
           of weak coupling perturbation theory of the action.
\item{(C)} For topologically nontrivial gauge backgrounds,
           it possesses exact zero modes satisfying the Atiyah-Singer
           index theorem.
\end{description}

Although Wilson's idea \cite{wilson74} of formulating gauge
theories on the spacetime lattice is the most successful
nonperturbative regularization for pure gauge fields,
putting massless Dirac fermions \cite{wilson75} on the lattice
has been a notorious problem for more than twenty years. The
resolution of the lattice fermion problem first appeared in the context
of the Domain-Wall fermion \cite{kaplan92}, and it motivated the
Overlap formalism \cite{rn95} which led to the construction
of Overlap-Dirac operator \cite{hn97:7} in 1997. We refer to
ref. \cite{tblum98:10} for a recent review of the Domain-Wall fermions,
and to ref. \cite{hn99:11} for a recent review of the Overlap.

However, if we consider a Weyl fermion interacting with a background
gauge field, then a completely satisfactory nonperturbative
regularization for chiral guage theories ( e.g., the standard model )
has not yet been presented up to this moment.

In the following, we will concentrate our discussions on the general
principles to construct chiral Dirac fermion operators on a finite
lattice, in particular, for vector-like gauge theories such as QCD.
With the constraints imposed by the Nielson-Ninomiya
no-go theorem\cite{no-go}, one can construct a gauge covariant Dirac
operator $ D $ on a finite lattice such that :
\begin{description}
\item[(i)] $ D $ breaks the chiral symmetry
         ( i.e., $ D \gm5 + \gm5 D = 0 $ ) at finite lattice spacing
         but recovers the chiral symmetry in the continuum limit $ a \to 0 $.
\item[(ii)] $ D $ is local. \newline
       ( $ | D(x,y) | \sim \exp( - | x - y | / l ) $ with $ l \sim a $; or
         $ D(x,y) = 0 $ for $ |x-y| > z $, where $ z $ is much less than
         the size of the lattice. )
\item[(iii)] In the free fermion limit, $ D $ is free of species doublings.
             \newline
         ( The free fermion propagator $ D^{-1}(p) $ has only one simple
           pole at the origin $ p = 0 $ in the Brillouin zone. )
\item[(iv)] In the free fermion limit, $ D $ has correct continuum behavior.
            \newline
         ( In the limit $ a \to 0 $, $ D(p) \sim i \gamma_\mu p_\mu $
           around $ p = 0 $. )
\end{description}
However, one cannot push the property {\bf (i)} any further,
while maintaining properties {\bf (ii)-(iv)}.
For example, if $ D $ is chirally symmetric at finite lattice spacing,
then it must violate at least one of the three properties {\bf (ii)-(iv)}.
We note that these four properties {\bf (i)-(iv)} form the necessary
conditions to meet the requirements {\bf (A)-(C)}, however, they
are not sufficient to guarantee that {\bf (C)} will be satisfied.

An example satisfying {\bf (i)-(iv)} is the standard
Wilson-Dirac fermion operator\cite{wilson75}
\bea
\label{eq:Dw}
D_w = \gamma_\mu t_\mu + w
\eea
where
\bea
\label{eq:tmu}
t_\mu (x,y) = \frac{1}{2} [   U_{\mu}(x) \delta_{x+\hat\mu,y}
                       - U_{\mu}^{\dagger}(y) \delta_{x-\hat\mu,y} ] \ ,
\eea
\beq
\label{eq:sigma}
\sigma_\mu \sigma_\nu^{\dagger} + \sigma_{\nu} \sigma_\mu^{\dagger} =
2 \delta_{\mu \nu} \ ,
\eeq
\bea
\label{eq:gamma}
\gamma_\mu &=& \left( \begin{array}{cc}
                            0                &  \sigma_\mu    \\
                    \sigma_\mu^{\dagger}     &       0
                    \end{array}  \right)
\eea
and $ w $ is the Wilson term
\bea
\label{eq:wilson}
w(x,y) =  \frac{a}{2} \sum_\mu \left[ 2 \delta_{x,y}
                     - U_{\mu}(x) \delta_{x+\hat\mu,y}
                     - U_{\mu}^{\dagger}(y) \delta_{x-\hat\mu,y} \right] \ .
\eea
The color, flavor and spinor indices have been suppressed in (\ref{eq:Dw}).
The first term on the r.h.s. of (\ref{eq:Dw}) is the naive fermion
operator which is chirally symmetric at any lattice spacing
and satisfies properties {\bf (ii)} and {\bf (iv)} but
violates {\bf (iii)} since it has $ 2^{d} - 1 $ fermion doubled modes.
The purpose of the Wilson term $ w $ is to give each doubled mode
a mass of $ \sim 1/a $ such that in the continuum limit ( $ a \to 0 $ ),
each doubled mode becomes infinitely heavy and decouples from the fermion
propagator. However, the introduction of the Wilson term has serious
drawbacks. It causes $ O(a) $ artifacts and also leads to the notorious
problems such as vector current renormalization, additive fermion mass
renormalization, and mixings between operators in different
chiral representations.

During the last two years, it has become clear that the proper
way to break the chiral symmetry of $ D $ at finite lattice spacing
is to conform with the Ginsparg-Wilson relation \cite{gwr}
\bea
\label{eq:gwr}
D \gm5 + \gm5 D = 2 a D R \gamma_5 D
\eea
where $ R $ is a positive definite hermitian operator which
is local in the position space and trivial in the Dirac space.
Then the generalized chiral symmetry (\ref{eq:gwr}) can
ensure that the theory is free of above mentioned
problems of the Wilson-Dirac operator \cite{ph98:2}.

The general solution to the Ginsparg-Wilson relation can be
written as \cite{twc98:6a}
\bea
\label{eq:gen_sol}
D = D_c ( \Id + a R D_c )^{-1} = ( \Id + a D_c R )^{-1} D_c
\eea
where $ D_c $ is any chirally symmetric ( $ D_c \gm5 + \gm5 D_c = 0 $ )
Dirac operator which must violate at least one of the three properties
{\bf (ii)}-{\bf (iv)} above.
Now we must require $ D_c $ to satisfy {\bf (iii)} and {\bf (iv)},
but violate {\bf (ii)} (  i.e, $ D_c $ is nonlocal ), since
(\ref{eq:gen_sol}) can transform the nonlocal $ D_c $ into a local $ D $
on a finite lattice for $ R = r \Id $
with $ r $ in the proper range \cite{twc98:9a,twc99:10},
while the properties {\bf (iii)} and {\bf (iv)} are preserved.
Moreover, the zero modes and the index of $ D_c $ are invariant under
the transformation \cite{twc98:6a,twc98:9a}.
That is, a zero mode of $ D_c $ is also a zero mode of $ D $
and vice versa, hence,
\beq
\label{eq:npm}
 n_{+} ( D_c ) = n_{+} ( D ), \hspace{4mm} n_{-} ( D_c ) = n_{-} ( D ),
\eeq
\beq
\label{eq:index}
\mbox{index}(D_c) = n_{-}(D_c) - n_{+}(D_c) =
 n_{-}(D) - n_{+}(D) = \mbox{index}(D) \ .
\eeq

Since the massless Dirac fermion operator in continuum is
antihermitan, we also require that $ D_c $ is antihermitian
( $ D_c^{\dagger} = -D_c $ ) even at finite lattice spacing.
Then the chiral symmetry of $ D_c $ together with its
antihermiticity implies that $ D_c $ satisfies the
$\gamma_5$-hermiticity
\bea
\label{eq:hermit}
 D_c^{\dagger} = \gm5 D_c \gm5 \ .
\eea
This implies that $ D $ in the general solution (\ref{eq:gen_sol})
also satisfies the $\gm5$-hermiticity
\bea
\label{eq:hermit_D}
D^{\dagger} = \gm5 D \gm5 \ .
\eea
Then the eigenvalues of $ D $ are either real or come in
complex conjugate pairs.
Furthermore, from (\ref{eq:gen_sol}), since $ R $ is a positive definite
hermitian operator, the lower bound of real eigenvalues of $ D $
is zero, thus $ \det(D) $ is real and nonnegative, and is amenable
to Hybrid Monte Carlo simulation for dynamical fermions with {\it any}
number of fermion flavors.

In particular, for $ R = r \Id $ with $ r > 0 $, the
analysis [ Eqs. (24)-(41) ] in ref. \cite{twc98:4} goes through
with trivial modifications. The main results are :
\begin{description}
\item[($\alpha$)] The eigenvalues of $ D $ fall on a circle with center
at $ 1/2r $, and radius $ 1/2r $, and have the reflection symmetry with
respect to the real axis.
\item[($\beta$)] The real eigenmodes ( if any ) at $ 0 $ and $ 1/r $ have
                 definite chirality $ +1 $ or $ -1 $.
\item[($\gamma$)] The chirality of any complex eigenmodes is zero.
\item[($\delta$)] Total chirality of all eigenmodes must vanish. \newline
  $ \Tr ( \gamma_5 ) = \sum_{s} \phi_s^{\dagger} \gamma_5 \phi_s
                     = n_{+} - n_{-} + N_{+} - N_{-} = 0 $
\end{description}
where $ n_{+} ( n_{-} ) $ denotes the number of zero modes of
positive ( negative ) chirality, and $ N_{+} ( N_{-} ) $ the number of
$ 1/r $ modes of positive ( negative ) chirality.
From ($\delta$), we immediately see that {\it any zero mode must be
accompanied by a real $ 1/r $ mode with opposite chirality}, and
the index of $ D $ is
\bea
\label{eq:index_D}
\mbox{index}(D) \equiv n_{-} - n_{+} = - ( N_{-} - N_{+} ) \ .
\eea

Now the central problem is to construct the chirally symmetric
$ D_c $ which is nonlocal, and satisfies {\bf (iii)}, {\bf (iv)},
and (\ref{eq:hermit}). Furthermore we also require that
$ D_c $ is topologically proper ( i.e., satisfying
the Atiyah-Singer index theorem ) for any prescribed smooth
gauge background. These constitute the necessary requirements
\cite{twc98:9a} for $ D_c $ to enter (\ref{eq:gen_sol})
such that $ D $ could provide a nonperturbative regularization
for a massless Dirac fermion interacting with a background
gauge field. Explicitly, these necessary requirements are :
\begin{description}
\item[(a)] $ D_c $ is antihermitian
           ( hence $\gamma_5$-hermitian )
           and it agrees with $ \gamma_\mu ( \partial_\mu + i g A_\mu ) $
           in the classical continuum limit.
\item[(b)] $ D_c $ is free of species doubling.
\item[(c)] $ D_c $ is nonlocal.
\item[(d)] $ D_c $ is well defined in topologically trivial background
           gauge field.
\item[(e)] $ D_c $ has zero modes as well as simple poles
           in topologically non-trivial background gauge fields
           ( each zero mode of $ D_c $ must be accompanied by
             a simple pole of $ D_c $ ).
           Furthermore, the zero modes of $ D_c $ satisfy the
           Atiyah-Singer index theorem for any prescribed smooth
           gauge background.
\end{description}

The general solution of $ D_c $ satisfying these requirements
had been investigated in ref. \cite{twc99:8} and was reported
by the author at Chiral'99. However, in general, given any lattice
Dirac operator $ D $, there exists a transformation $ \CT(R_c) $
for $ D $ such that the transformed Dirac operator $ D_c = \CT(R_c) [D] $
is chirally symmetric \cite{twc99:6,twc99:12a}. Therefore, for the sake
of completeness, the transformation $ \CT(R_c) $ is outlined in section 2.
Then the construction in ref. \cite{twc99:8} is reviewed in section 3.
Concluding remarks are briefly outlined in section 4.

\section{ A transformation for lattice Dirac operators }

Given any lattice Dirac operator $ D $, in general, there are many
different ways to construct a chirally symmetric $ D_c $ out of $ D $.
For example, we can construct
\bea
\label{eq:Ds}
D_s = \frac{1}{2} ( D - \gamma_5 D \gamma_5 )
\eea
which is chirally symmetric ( $ D_s \gamma_5 + \gamma_5 D_s = 0 $ ).
However, this transformation does not necessarily preserve the
property {\bf (iii)}. For example, if we apply this transformation
to the Wilson-Dirac operator (\ref{eq:Dw}),
we obtain $ D_s = \gamma_\mu t_\mu $, the naive
fermion operator which suffers from the species doublings. Although
$ D_w $ is free of species doublings in the continuum limit, the
transformation (\ref{eq:Ds}) cannot preserve this property since
$ D_s $ is local. Therefore, we need a transformaton which preserves
the properties {\bf (iii)}, {\bf (iv)}, (\ref{eq:hermit_D})
and (\ref{eq:npm}), but exchanges the locality of $ D $ for
its chiral symmetry at finite lattice spacing.
The transformation \cite{twc99:6, twc99:12a}
\bea
\label{eq:tit}
\CT(R) :  \hspace{4mm} D \to D'= \CT(R)[D] \equiv D ( \Id + R D )^{-1}
                               = ( \Id + D R )^{-1} D
\eea
which generalizes (\ref{eq:gen_sol}) can serve our purposes.
Evidently, the set of these transformations, $ \{ \CT(R) \} $,
form an abelian group with group parameter space
$\{ R \}$ \cite{twc99:6,twc99:12a}.

For any $ D $ satisfying $\gamma_5$-hermiticity, (\ref{eq:hermit_D}),
there exists a hermitian $ R_c $,
\bea
\label{eq:Rc}
R_c = - \frac{1}{2} ( D^{-1} + \gamma_5 D^{-1} \gamma_5 )
\eea
such that
\bea
\label{eq:D5D}
D_c = \CT(R_c) [D]
    = 2 \gm5 D ( \gm5 D - D \gm5 )^{-1} D
\eea
is chirally symmetric ( $ \gamma_5 D_c + D_c \gamma_5 = 0 $ ) and
antihermitian ( $ D_c^{\dagger} = -D_c $ ).

It is evident that for any two lattice Dirac operators $ D^{(1)} $ and
$ D^{(2)} $ satisfying (\ref{eq:hermit_D}),
their corresponding chiral limits obtained from (\ref{eq:D5D}),
say, $ D_c^{(1)} $ and $ D_c^{(2)} $, are in general different.
However, they are related by the transformation
\bea
\label{eq:DcT}
D_c^{(1)} = \sum_{i} T_i D_c^{(2)} T_{i}^{\dagger}
\eea
where each $ T_i $ commutes with $ \gm5 $. In general,
given any two $ D_c^{(1)} $ and $ D_c^{(2)} $,
it is a nontrivial task to obtain all $ T_i $ in (\ref{eq:DcT}).

For the Wilson-Dirac operator (\ref{eq:Dw}),
the transformation (\ref{eq:D5D}) gives
\bea
\label{eq:Dtwc}
D_c = \gamma_\mu t_\mu - w \ ( \gamma_\mu t_\mu )^{-1} \ w  \ .
\eea
It is antihermitian, nonlocal ( due to the second term ),
and satisfies {\bf (i), (iii)} and {\bf (iv)}.
Then we can substitute (\ref{eq:Dtwc}) into (\ref{eq:gen_sol})
with $ R = r \Id $ to obtain a GW Dirac operator
\bea
\label{eq:Dw_GW}
D = \left[ \begin{array}{cc}
 r \ C^{\dagger} C ( \Id  + r^2 C^{\dagger} C )^{-1}  &
    - C^{\dagger} ( \Id + r^2 C C^{\dagger} )^{-1}      \\
 C ( \Id + r^2 C^{\dagger} C )^{-1}  &
 r \ C C^{\dagger} ( \Id + r^2 C C^{\dagger} )^{-1}
            \end{array}      \right]
\eea
where
\bea
\label{eq:C}
C = ( \sigma^{\dagger}_\mu t_\mu ) - w ( \sigma_\mu t_\mu )^{-1} w \ .
\eea

On a finite lattice, the GW Dirac operator (\ref{eq:Dw_GW}) can be
constructed to be local but not highly peaked in the diagonal elements
if the value of $ r $ is within a proper range \cite{twc99:10}. In other
words, if $ r $ is too small, then $ D $ must be nonlocal since $ D $
is close to $ D_c $. On the other hand, if $ r $ is too large,
then $ D $ becomes highly peaked in the diagonal elements.
In these two extreme cases,
$ D $ does not respond properly to the background gauge field ( e.g.,
the anomaly function of $ D $ does not agree with the Chern-Pontryagin
density of the prescribed gauge background ). In the limit the number
of sites in each dimension goes to infinity, the proper range
of $ r $ extends to $ ( 0, \infty ) $.

The anomaly function of $ D $ (\ref{eq:Dw_GW}) can be written
\bea
\label{eq:anxc}
\anxc = 2 \ \tr [   ( \Id + r^2 C C^{\dagger} )^{-1}
                - ( \Id + r^2 C^{\dagger} C )^{-1}   ] (x,x)
\eea
where $ \tr $ denotes the trace over the color, flavor and spinor space.

It is instructive to compare the anomaly function of the Wilson-Dirac
operator $ D_w $ (\ref{eq:Dw}) to that of the GW Wilson-Dirac operator
$ D $ (\ref{eq:Dw_GW}), in a topologically trivial background gauge field,
on a finite lattice, as shown in Fig. 1 in ref. \cite{twc99:11}
and Fig. 1 in ref. \cite{twc99:12a}, respectively. The anomaly function
of $ D_w $ is very different from the Chern-Pontryagin
density $ \rho(x) $ on a $ 12 \times 12 $ torus, while that of the
GW Dirac-Wilson operator $ D $ is
in good agreement with $ \rho(x) $ at each site.
This demonstrates that the transformation $ \CT( r + R_c ) $ plays
the important role in converting $ D_w $ into a GW Dirac operator
$ D = \CT( r + R_c ) [ D_w ] $ which is free of $ O(a) $
lattice artifacts, thus $ D $ can reproduce the correct anomaly function
even on a {\it finite} lattice provided that the local fluctuations
of the background gauge field are not too violent.

However, a lattice Dirac operator satisfying properties
{\bf (i)-(iv)}, (\ref{eq:hermit_D}) and the GW relation (\ref{eq:gwr})
does not guarantee that it has the correct anomaly function in topologically
nontrivial gauge backgrounds. This can be seen as follows.
It is well known that the sum of the axial anomaly
over all sites of a finite lattice is equal to the index of the lattice
Dirac operator. Therefore {\it the necessary condition for a lattice
Dirac operator to have the correct anomaly function is that
it possesses exact zero modes satisfying the Atiyah-Singer index theorem}
i.e., $ n_{-} - n_{+} = Q $. However, not every lattice Dirac operator
has exact zero modes in topologically nontrivial gauge backgrounds.
For example, the Wilson-Dirac operator, $ D_w $ in (\ref{eq:Dw}), it
does not have exact zero modes in topologically nontrivial
sectors. Then, according to (\ref{eq:npm}), $ D = \CT( r + R_c ) [ D_w ] $
also does not have any exact zero modes, even though $ D $ satisfies the GW
relation (\ref{eq:gwr}), and properties {\bf (i)-(iv)} and (\ref{eq:hermit_D}).
This implies that the anomaly function of $ D_w $ or
$ D = \CT( r + R_c ) [ D_w ] $ does not agree with the
Chern-Pontryagin density for topologically
nontrivial gauge fields on {\it any} finite lattices.
Consequently, {\it the disagreement must persist in the
continuum limit $ a \to 0 $}.

This provides an example to illustrate that
any lattice Dirac operator $ D $ must possess a nonperturbative
attribute, the topological characteristics, $ c[D] $ \cite{twc98:9a,twc99:11}.
In general, $ c[D] $ is a rational number, a functional of $ D $ and the
gauge configuration. In a gauge background with nonzero topological
charge $ Q $, the index of a lattice Dirac operator is
\bea
\label{eq:cD}
n_{-} - n_{+} = c[D] \ Q \ .
\eea
In the case of $ D_w $ and $ D = \CT( r + R_c ) [ D_w ] $ (\ref{eq:Dw_GW}),
$ n_{-} = n_{+} = 0 $, thus the topological characteristics
$ c[D_w] = c[D] = 0 $. For the vacuum
sector with nonzero topological charge density ( i.e., $ Q = 0 $ but
$ \rho(x) \ne 0 $ ), $ c[D] $ cannot be defined by (\ref{eq:cD}), however,
it can be defined unambiguously \cite{twc99:11} in the context of the
anomaly function.

We note in passing that for the Overlap-Dirac operator
$ D_o = \Id + V $ \cite{hn97:7}, the transformation (\ref{eq:D5D})
gives $ D_c = 2 ( \Id + V ) ( \Id - V )^{-1} $, as expected.
Substituting $ D_c $ into (\ref{eq:gen_sol}) with $ R = r \Id $ ,
we obtain the generalized Overlap-Dirac operator
\beq
\label{eq:Do_GW}
D = 2 ( \Id + V ) [ \Id - V + 2 r ( \Id + V ) ]^{-1} \ .
\eeq

\section{ A construction of $ D_c $ }

Suppose that we do not have any lattice Dirac operator
to begin with. It is still possible for us to construct
$ D_c $ \cite{twc99:8} according to the necessary
requirements {\bf (a)-(e)} listed in section 1.
Since $ D_c $ is antihermitian, there exists one to one
correspondence between $ D_c $ and a unitary operator $ V $,
\bea
D_c = (\Id + V )(\Id - V )^{-1}, \hspace{4mm}
V = (D_c - \Id)( D_c + \Id)^{-1}.
\label{eq:VDc}
\eea
where $ V $ also satisfies the $\gamma_5$-hermiticity,
$ \gamma_5 V \gamma_5 = V^{\dagger} $.
Then the unitary operator $ V $ can be expressed in terms of
a hermitian operator $ h $,
\bea
\label{eq:V5h}
V = \gamma_5 h = \left( \begin{array}{cc}
                         \Id  &    0    \\
                           0  & -\Id
                        \end{array}      \right)
                  \left( \begin{array}{cc}
                          h_1            &  h_2    \\
                          h_2^{\dagger}  &  h_3
                        \end{array}      \right)
               =  \left( \begin{array}{cc}
                          h_1            &  h_2    \\
                         -h_2^{\dagger}  & -h_3
                        \end{array}      \right)
\eea
where $ h_1^{\dagger} = h_1 $ and $ h_3^{\dagger} = h_3 $.
Using the unitarity condition $ V^{\dagger} V = \Id $,
we have $ h^2 = \Id $,
\bea
\label{eq:h^2}
h^2 = \left( \begin{array}{cc}
     h_1^2 + h_2 h_2^{\dagger}              &   h_1 h_2 + h_2 h_3    \\
     h_2^{\dagger} h_1 + h_3 h_2^{\dagger}  &   h_2^{\dagger} h_2 + h_3^2
            \end{array}      \right)
    = \left( \begin{array}{cc}
              \Id          &  0    \\
                0          &  \Id
             \end{array}      \right)
\eea
Then we obtain
\bea
D_c
\label{eq:Dc}
\equiv \left[ \begin{array}{cc}
                     0          &   D_R  \\
                     D_L        &   0
              \end{array}                           \right]
=  \left[ \begin{array}{cc}
                     0                 &  (\Id - h_1 )^{-1} h_2  \\
     -h_2^{-1} (\Id + h_1)             &   0
              \end{array}                           \right]
\eea
where $ D_L = - D_R^{\dagger} $.
The general solution to Eq. (\ref{eq:h^2}) can be written as
\bea
\label{eq:h1}
h_1 &=& \pm \frac{1}{ \sqrt{ \Id + b b^{\dagger} } } \\
\label{eq:h2}
h_2 &=& \frac{1}{ \sqrt{ \Id + b b^{\dagger} } } \ b \ e^{ i \theta }
\eea
where $e^{i\theta}$ is an arbitrary phase, and $b$ is any operator.
In the following we will restrict $ \theta $ to zero,
and also pick the minus sign for $ h_1 $. Then the general solution
for $ D_c $ can be written in the following form
\bea
\label{eq:DL}
D_L &=& b^{-1} \left[ \Id - \sqrt{\Id + b b^{\dagger}} \ \right] \\
\label{eq:DR}
D_R &=& \left[ \Id + \sqrt{\Id + b b^{\dagger}} \ \right]^{-1}  b
\eea
Due to the presence of the square root in (\ref{eq:DL}) and (\ref{eq:DR}),
$ D_c $ is nonlocal for nontrivial $ b $, thus the requirement {\bf (b)}
is satisfied. In the following, we outline a construction
of $ b $ which satisfies constraints {\bf (a)-(d)} but not {\bf (e)}.

In order to have $ D_c $ satisfy the constraint {\bf (a)}, we first try
\bea
\label{eq:b}
b = w^{-1} \sum_\mu \sigma_\mu t_\mu
\eea
where $ t_\mu $ and $ \sigma_\mu $ are defined in (\ref{eq:tmu})
and (\ref{eq:sigma}) respectively, and $ w $ is a non-singular
hermitian operator which is trivial in the Dirac space and goes to a
constant in the classical continuum limit\footnote{ The $ w $ operator
in this section is different from the Wilson term in Eq. (\ref{eq:wilson}). }.
Then Eqs. (\ref{eq:DL}) and (\ref{eq:DR}) become
\bea
\label{eq:DLw}
D_L &=& ( \sigma \cdot t )^{-1}
        \left[ w - \sqrt{ w^2 } \sqrt{ 1 + w^{-1} \ t^2 \ w^{-1} } \ \right] \\
\label{eq:DRw}
D_R &=& - D_L^{\dagger}
\eea
where
\bea
\label{eq:sigma_t}
\sigma \cdot t &=& \sum_\mu \sigma_\mu t_\mu   \\
\label{eq:t2}
t^2 &=& - ( \sigma \cdot t ) ( \sigma^{\dagger} \cdot t )
\eea
However, $ t_\mu $ suffers from species doublings. Now our task
is to construct a hermitian operator $ w $ such that the doubled
modes are completely decoupled from the fermion
propagator $ D_c^{-1} $, even at finite lattice spacing.

As discussed in ref. \cite{twc99:8}, we want to construct a hermitian $ w $
such that in the free fermion limit, it satisfies the following condition,
\bea
\label{eq:wp}
w(p) = \left\{  \begin{array}{ll}
  > 0   &  \mbox{ for the primary mode at $ p = 0 $  }     \\
  < 0   &  \mbox{ for the doubled modes at
   $ p \in \otimes_{\mu} \{ 0, \pi/a \} \backslash \{ p=0 \} $.   }  \\
                    \end{array}
              \right.
\eea
A simplest solution to (\ref{eq:wp}) is
\bea
\label{eq:wpc}
w(p) = c - 2 \sum_\mu  \sin^2 ( p_\mu a / 2), \hspace{4mm} c \in ( 0, 2 )
\eea
Note that the role of $ w $ in the general solution of $ D_c $ is quite
different from the Wilson term (\ref{eq:wilson}) in the Wilson-Dirac
operator (\ref{eq:Dw}). In the general solution of $ D_c $,
the chiral symmetry is always preserved,
and {\it the role of $ w $ is to suppress the doubled modes completely
at finite lattice spacing}, while in the Wilson-Dirac operator,
the Wilson term breaks the chiral symmetry explicitly and gives a mass
of order $ a^{-1} $ to the doubled modes such that they can be
decoupled in the continuum limit ( $ a \to 0 $ ).
After the gauge links are restored, $ w $ in the position space becomes
\bea
\label{eq:wxy}
w(x,y) = c \ \delta_{x,y} - \frac{1}{2} \sum_\mu \left[ 2 \delta_{x,y}
                       - U_{\mu}(x) \delta_{x+\hat\mu,y}
                       - U_{\mu}^{\dagger}(y) \delta_{x-\hat\mu,y} \right]
\eea
This is one of the simplest solution of $ w $ satisfying the
requirement (\ref{eq:wp}) in the free fermion limit.
Certainly, there exists other solutions to (\ref{eq:wp}).

Substituting the $ D_c $ [ Eqs. (\ref{eq:DLw})-(\ref{eq:DRw})
with $ t_\mu $ in (\ref{eq:tmu}), and $ w $ in (\ref{eq:wxy})
with $ c = 1 $, and the normalization constant $ 1/2 $ ]
into (\ref{eq:gen_sol}) with $ R = r \Id $,
we obtain the GW Dirac operator
\bea
\label{eq:twc}
D = \left[ \begin{array}{cc}
 r ( B B^{\dagger} + r^2 )^{-1}  &  - B ( B^{\dagger} B + r^2 )^{-1}      \\
 B^{\dagger} ( B B^{\dagger} + r^2 )^{-1}  &  r ( B^{\dagger} B + r^2 )^{-1}
            \end{array}      \right]
\eea
where
\bea
\label{eq:Bwt}
B &=& \frac{1}{2} \left( w - \sqrt{w^2} \sqrt{\Id + w^{-1} \ t^2 \ w^{-1} } \
                  \right)^{-1} \ ( \sigma \cdot t )
\eea
The anomaly function of $ D $ is
\bea
\label{eq:anx}
\anxc = 2 r^2 \ \tr \left[ ( B B^{\dagger} + r^2 )^{-1} -
                           ( B^{\dagger} B + r^2 )^{-1} \right] (x,x)
\eea
where $ \tr $ stands for the trace over the color, flavor and spinor space.

Since the $ D_c $ of (\ref{eq:twc})
in the free fermion limit is free of species doubling and has the
correct continuum behavior, the perturbation calculation in
ref. \cite{twc99:1} showed that $ D = D_c ( \Id + r D_c )^{-1} $
has the correct chiral anomaly in topologically trivial gauge backgrounds.
This has been verified explicitly on a finite lattice.
An example is shown in Fig. 2 of ref. \cite{twc99:11}.

However, for topologically nontrivial sectors, the anomaly function of $ D $
also depends on the topological characteristics, $ c[D] $,
which is a non-perturbative attribute of $ D $.
Since $ D $ does not have any exact zero modes in nontrivial sectors,
the anomaly function of $ D $ does not agree with the
Chern-Pontryagin density for nontrivial gauge backgrounds.
Now we have two examples of GW Dirac operators, (\ref{eq:Dw_GW}) and
(\ref{eq:twc}), which have the correct axial anomaly in the vacuum sector
but not in the nontrivial sectors.

In general, if $ D_c $ is well defined in nontrivial sectors ( i.e.,
not satisfying {\bf (e) } ),
then $ \mbox{index}[D_c] = \mbox{index}[D] = 0 $ \cite{twc98:6a,twc98:9a}
( i.e., $ c[D] = c[D_c] = 0 $, $ D $ and $ D_c $ are {\it topologically
trivial} ). On the other hand, if $ D_c $ has zeros and poles, then
$ c[D] \ne 0 $, however, it does not necessarily imply that $ c[D] = 1 $.
Sometimes $ c[D] $ may even become a fraction
( see Table 4 in ref. \cite{twc99:6} ).
The important point is that in general we cannot predict the value of
$ c[D] $ in the topologically nontrivial sectors solely based on the
properties of $ D $ in the free fermion limit or in the vacuum sector.

\section{ Summary and discussions }

The necessary conditions for a lattice Dirac operator $ D $
to provide a nonperturbative regularization for
massless fermion interacting with a background gauge field are
{\bf (A)-(C)} listed in section 1.
The conditions {\bf (A)} and {\bf (B)} can be satisfied if $ D $ fulfils
{\bf (i)-(iv)} and (\ref{eq:hermit_D}), and the chiral symmetry is broken
according to the Ginsparg-Wilson relation (\ref{eq:gwr}).
Since any GW Dirac operator can be represented by the transformation
(\ref{eq:gen_sol}) [ preserving {\bf (iii), (iv)}, (\ref{eq:hermit_D})
and (\ref{eq:npm}) ] on a chirally symmetric $ D_c $,
then the conditions for $ D $ become
the necessary requirements {\bf (a)-(e)} for $ D_c $.
It should be emphasized that the requirement {\bf (C)} for $ D $
( or the requirement {\bf (e)} for $ D_c $ )
cannot be expressed in terms of any conditions in the free fermion
limit or in the vacuum sector.
This naturally leads to the concept of topological characteristics
\cite{twc98:9a, twc99:11} associated with each lattice Dirac
operator $ D $. Although we have no problems to construct $ D $ to
satisfy {\bf (i)-(iv)}, (\ref{eq:hermit_D}) and (\ref{eq:gwr}),
it remains an interesting question whether one can construct a
topologically proper $ D $ on
a four dimensional lattice without square root operations.

If one has a lattice Dirac operator $ D $ satisfying
{\bf (i)-(iv)}, (\ref{eq:hermit_D}) and (C),
then one can transform $ D $ into a GW Dirac operator
through the transformation $ D' = \CT( r + R_c )[D] $ which not only
preserves all essential physics of $ D $ but also can restrict
the quantum corrections to behave properly such that $ D $
is free of $ O(a) $ lattice artifacts, additive mass
renormalization, etc.

Recently it has been shown that the effective four-dimensional
action ( $ N_s \to \infty $ ) of the Domain Wall fermion is local
and satisfying the GW relation, for gauge fields with small
field strength \cite{kiku99:12}. However, it is
possible to transform the Domain-Wall fermion operator into
a local GW Dirac operator even at finite $ N_s $ \cite{twc99:12a},
then the anomalous effects
( e.g., the residual pion mass in Domain-Wall QCD )
due to chiral symmetry violations at finite $ N_s $ could be
suppressed even at moderate $ N_s $. Further investigations are
required before one can tell whether this scenario can be realized or not.

\bigskip
\bigskip


\flushpar
{\bf Acknowledgement }
\bigskip

\noindent
I would like to thank all participants of Chiral'99, for their
inspiring talks, interesting questions, and enlightening discussions.
This work was supported by the National Science Council, R.O.C.
under the grant number NSC89-2112-M002-017.

\bigskip
\bigskip


\end{document}